\documentclass[12pt]{article}
\usepackage[T1]{fontenc}
\usepackage[latin1]{inputenc}
\usepackage{
fixltx2e,
ellipsis,
ragged2e,
marginnote}
\usepackage[tracking=true]{microtype} 
\usepackage{lmodern}
\usepackage{amsmath}           
\usepackage{amssymb}
\usepackage{booktabs}
\usepackage{multirow}
\usepackage{icomma}            
\usepackage[thinqspace, squaren, textstyle]{SIunits}           
\usepackage[english]{babel}    
\usepackage{fancyhdr}         
\usepackage{color}            
\usepackage{colordvi}         
\usepackage{cancel}
\usepackage{colortbl}
\usepackage[bookmarksopen=true, hidelinks]{hyperref} 
\usepackage{braket}
\usepackage{caption, threeparttable}
\usepackage{graphicx}					
\usepackage[format=plain,labelsep=period,labelfont=bf]{caption}
\usepackage{times}
\usepackage{url} 
\usepackage[a4paper,margin=2cm]{geometry}
\usepackage{authblk}
\usepackage[sort&compress]{natbib}
\bibpunct{}{}{,}{s}{}{\textsuperscript{,}}

\date{}
\title{\textbf{Complete Coherent Control of a Quantum Dot Strongly Coupled to a Nanocavity}}

\author[1,*,+]{Constantin~Dory}
\author[1,+]{Kevin~A.~Fischer}
\author[1,+]{Kai~Müller}
\author[1]{Konstantinos~G.~Lagoudakis}
\author[1]{Tomas~Sarmiento}
\author[1]{Armand Rundquist}
\author[1]{Jingyuan~L.~Zhang}
\author[1]{Yousif Kelaita}
\author[1]{Jelena Vu\v{c}kovi\'c}

\affil[1]{E. L. Ginzton Laboratory, Stanford University, Stanford, California 94305, USA}
\affil[*]{Correspondence to cdory@stanford.edu}
\affil[+]{These authors contributed equally.}

\begin{document}
	\maketitle

\begin{abstract}
Strongly coupled quantum dot-cavity systems provide a non-linear configuration of hybridized light-matter states with promising quantum-optical applications. Here, we investigate the coherent interaction between strong laser pulses and quantum dot-cavity polaritons.  Resonant excitation of polaritonic states and their interaction with phonons allow us to observe coherent Rabi oscillations and Ramsey fringes. Furthermore, we demonstrate complete coherent control of a quantum dot-photonic crystal cavity based quantum-bit. By controlling the excitation power and phase in a two-pulse excitation scheme we achieve access to the full Bloch sphere. Quantum-optical simulations are in good agreement with our experiments and provide insight into the decoherence mechanisms.

\end{abstract}

\flushbottom

\thispagestyle{empty}
\noindent

\section*{Introduction}

The rapid technological development of classical computing will soon reach fundamental limitations resulting from device miniaturization. However, the quantum regime and integration of optics on existing computing platforms offer a wide range of possibilities for overcoming the obstacles of Moore's law or charge carrier mobility.

First approaches to develop quantum technologies were made in atomic physics \cite{Monroe2002}, while real-world applications are more likely to be realized in solid state quantum systems \cite{Vuckovic2003, Jelezko2006, Michler2009, MasBalleste2011}. Self-assembled quantum dots (QDs) are particularly attractive due to their optical addressability, their narrow linewidths and ease of integration into optoelectronic devices. To make use of these promising characteristics, complete coherent control of the quantum device is a necessity. For QDs this has already been widely explored in excitons \cite{Stievater2001, Zrenner2002, Ramsay2010a}. Yet, the operational-range and the physical properties of QDs are limited, in particular low emission rates impede their application as non-classical light sources in photonic devices. Furthermore, the strength of their coupling to light poses challenges for the on-chip realization of quantum networks.   

In contrast, strongly coupled QD-cavity systems offer highly efficient out-coupling allowing for on-chip integration \cite{Englund2005, Brossard2010, Reinhard2011, Kim2013}. Moreover, they have proven to be extremely versatile, since the hybridization of electromagnetic waves and matter forms polaritons, yielding rich physical characteristics. In particular, it is possible to control the spontaneous emission of a QD in an optical cavity\cite{Lodahl2004}, while the interaction of excitons and phonons\cite{Wilson-Rae2001} allows for applications such as indistinguishable photon generation\cite{Muller2015c}. Future applications can profit from the systems' interesting and especially useful physics far beyond that offered by QDs alone. Examples include high-fidelity photon-blockade \cite{Muller2015a}, non-classical light generation \cite{Muller2015}, single spin-photon interfaces enabling compact on-chip quantum circuits \cite{Sun2015} and dynamic cavity frequency tuning with surface acoustic waves \cite{Blattmann2014}. One crucial element that is needed for successful on-chip integration of strongly coupled systems is coherent control.
  
As discussed theoretically \cite{Clark2007}, dissipation hindered the complete coherent control of QD-cavity systems so far. In this work, we overcome this dissipation obstacle and demonstrate complete coherent control of a QD-photonic crystal cavity system. In experiments supported by simulations, we map out the excitation power and phase-dependent emission from a polaritonic system and we demonstrate full access of the Bloch sphere.

\section*{Results}
\subsection*{Strongly coupled QD-photonic crystal cavity system}

A system composed of a single QD strongly coupled to an L3 photonic crystal cavity \cite{Akahane2003} is the focus of our investigations. This type of cavity offers very high quality factors as well as low mode volumes \cite{Arakawa2012}, which allows the light-matter interaction to reach the strong-coupling regime \cite{Hennessy2006}. Details on the sample fabrication can be found in the Methods section.

The physics of a strongly coupled system is described by the Jaynes-Cummings (JC) model \cite{Jaynes1963}, with the Hamiltonian $\mathcal{H}=\mathcal{H}_{C} + \mathcal{H}_{QD} + \mathcal{H}_{Int}$. The contributions are 
\begin{equation}
\mathcal{H}_{C} = \omega_{C} a^{\dagger} a,~~~\mathcal{H}_{QD}=\left(\omega_{C} + \Delta\right)\sigma^\dagger \sigma ~~~\text{and} ~~~\mathcal{H}_{Int}= g(a^\dagger\sigma+a\sigma^\dagger),
\label{eq:Hamiltonian}
\end{equation}
with the cavity field Hamiltonian $\mathcal{H}_{C}$, the QD Hamiltonian $\mathcal{H}_{QD}$ and the QD-cavity interaction Hamiltonian $\mathcal{H}_{Int}$, respectively. The cavity frequency is given by $\omega_{C}$, while $a$ is the cavity mode photon annihilation operator, $\sigma$ the QD's lowering operator, $\Delta$ the QD-cavity detuning and $g$ the coupling strength of the quantum emitter and the cavity. 

A full description of the system's complex eigenenergies is obtained by the addition of dissipation to the system and consists of a series of polariton rungs \cite{Laussy2014}:
\begin{equation}
	\mathcal{E}_\pm^n = n\omega_C + \frac{\Delta}{2}-i\frac{\left(2n-1\right)\kappa+\gamma}{4} \pm\sqrt{\left(\sqrt{n}g\right)^2+\left(-\frac{\Delta}{2}-i\frac{\kappa-\gamma}{4}\right)^2},
	\label{eq:PRXEnergy}
\end{equation}
where $\mathcal{E}_\pm^n$ represents the energy of the $n$th rung, while $\kappa$ and $\gamma$ are the cavity and the QD energy decay rates, respectively.   

\begin{figure}[ht]
	\centering
		\includegraphics{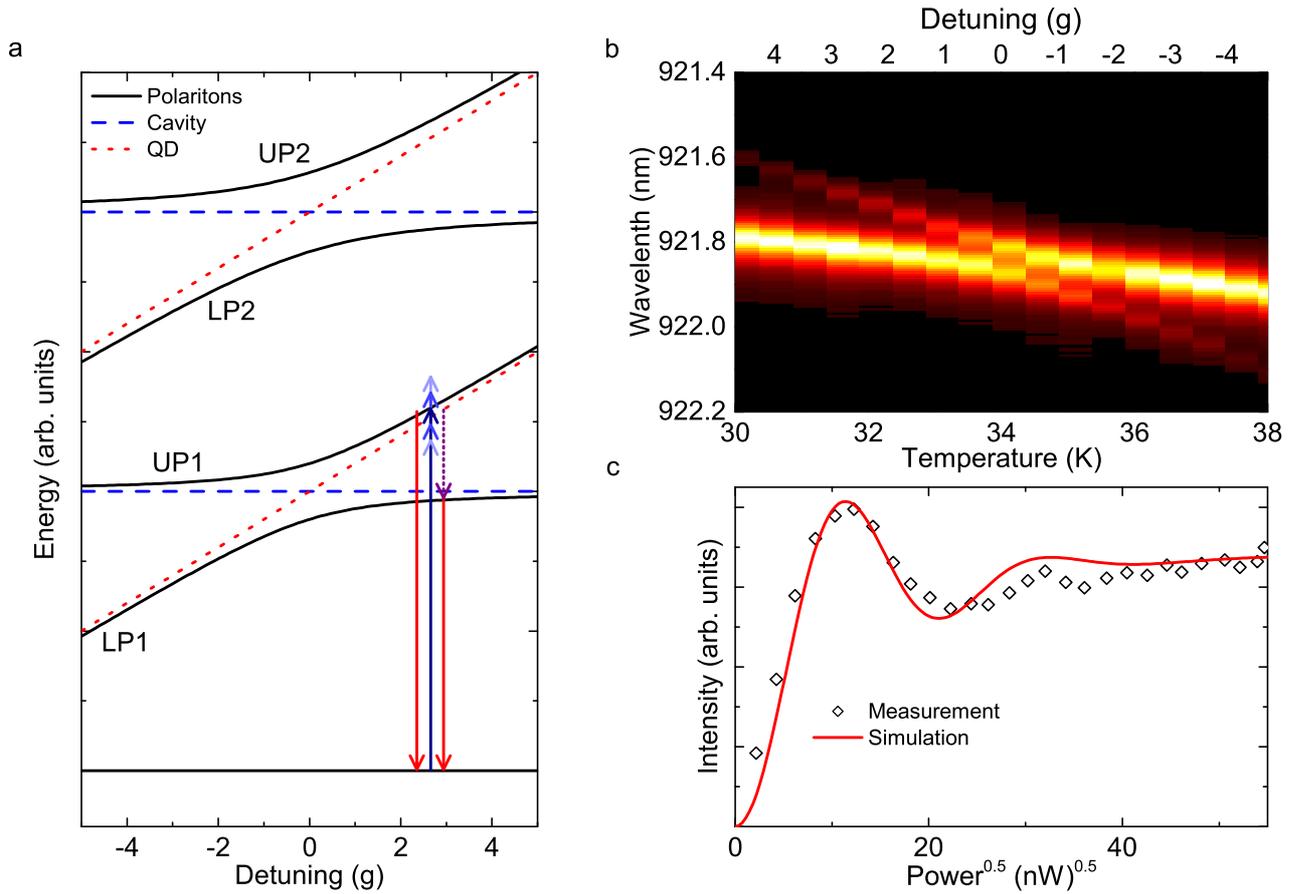}
	\caption{\textbf{Characteristics of a strongly coupled QD-photonic crystal cavity system.} (\textbf{a}) Energy level structure of a strongly coupled system resulting from the Jaynes-Cummings model (equation \ref{eq:PRXEnergy}). Resonant excitation of UP1 is illustrated by a dark blue arrow, while the spectral width of the excitation pulse is illustrated with a sequence of blue arrows. The purple arrow illustrates phonon-assisted population transfer between the polariton branches and the red arrows illustrate radiative recombination. (\textbf{b}) Measurement of the strongly coupled QD-cavity system emission acquired in cross-polarized reflectivity. The QD frequency is tuned in and out of the cavity resonance by changing the crystal lattice temperature. (\textbf{c}) Rabi oscillations for resonant excitation of UP1, while detecting emission from LP1. Quantum optical simulations to the data are shown as a red line.}
	\label{fig:General}
\end{figure}

The energy level structure resulting from equation \ref{eq:PRXEnergy}, called the JC ladder, is shown in figure \ref{fig:General}a. The polariton rungs are indexed by $n$ and each rung is divided into a lower polariton (LP$n$) and an upper polariton (UP$n$). Red dotted lines indicate the energy levels of the uncoupled quantum emitter and blue dashed lines the ones of the cavity. The polariton branches show anticrossings which are known as a convincing signature of strong coupling \cite{Reithmaier2004, Peter2005}.

Experimentally, the anti-crossing of the lowest energy rung can be measured using cross-polarized reflectivity \cite{Englund2007}. The result of a typical measurement is presented in figure \ref{fig:General}b. Here, the detuning between QD and cavity is controlled by changing the sample temperature \cite{Hennessy2006, Englund2007, Kaniber2008a}. Fitting this data with equation \ref{eq:PRXEnergy} reveals values of $g= 12.3 \cdot 2\pi$~GHz and $\kappa= 18.4 \cdot 2\pi$~GHz. The value of $\gamma$ cannot be fitted directly as it is much smaller than the other rates due to the highly dissipative character of solid-state systems. Typical values in literature are $\gamma \sim 0.01 \cdot 2\pi$~GHz \cite{Englund2007, Kaniber2008}. Importantly, compared to QDs in a bulk environment, the QD energy decay rate is further reduced due to Purcell suppression in the photonic bandgap. 

\subsection*{Coherent control of JC polaritons}
The coherent control of the population of exciton states in self-assembled QDs \cite{Zrenner2002, Gaudreau2011, Webster2013} or of quantum well-microcavity polaritons \cite{Dominici2014} is routinely performed. However, for polaritonic states of JC systems, the situation is more complicated due to the higher rungs of the JC ladder. In particular, for currently achievable system parameters and the QD in resonance with the cavity, a short-pulsed laser in resonance with the first rung would also be resonant with higher rungs of the ladder, preventing coherent control of a specific rung \cite{Muller2015a}. Moreover, the presence of the higher rungs results in strong coherent scattering of the excitation laser \cite{Fischer2015}, which would dominate over the signal of interest at excitation powers needed for coherent control.

We will now detail how we overcome both of these obstacles. To achieve a coherent interaction with a specific rung, the excitation laser needs to be spectrally narrow enough to prevent overlap with other rungs. At the same time, the pulse length needs to be shorter than the state lifetime to prevent re-excitation. At zero QD-cavity detuning all resonances of the JC ladder are very close. However, with increasing detuning between the QD and cavity, the energy difference between the two polariton branches within one rung becomes larger as the polaritons gradually evolve into the bare QD and cavity states \cite{Muller2015a}. Importantly, when exciting the QD-like polaritonic branch with increasing detuning, the overlap with higher rungs decreases \cite{Muller2015a}. Moreover, when the state becomes more QD-like, then the lifetime increases. This allows for a compromise between laser spectral width and pulse length to resonantly access the individual polariton branches (see figures \ref{fig:General}a and \ref{fig:General}b). 

Note that the polaritons are a superposition of the bare QD eigenstates and the bare cavity eigenstates. During the experiments we will focus on detunings of $\Delta= 6.75~g$ and $\Delta = 8.5~g$. At a detuning of $6.75~g$ we find UP1 to be approximately $97.9~\%$ QD-like and $2.1~\%$ cavity-like, while it is approximately $98.7~\%$ QD-like and $1.3~\%$ cavity-like at a detuning of $8.5~g$. Although the influence of the cavity seems to be insignificant, the QD-like polariton emits almost only through the cavity mode. This is due to the Purcell suppression of non-cavity modes and is true for detunings of up to approximately $60~g$\cite{Muller2015a}. In fact, the cavity completely determines the system's dynamics and the strong coupling leads to emission rates and an oscillator strength that is three orders of magnitude stronger than in the weak coupling regime. Thus, it is extremely difficult to observe the emission of a weakly coupled system in standard cross-polarized resonance fluorescence since the coherent scattering of the laser light dominates the signal.

We exploit the efficient exciton-phonon coupling to avoid the strong coherent scattering of the excitation laser dominating the resonance fluorescence signal. Investigations of the polariton-phonon interactions have led to a better understanding of the system dynamics \cite{Wilson-Rae2001, Ota2009}. Specifically, it was found that exciton-phonon coupling leads to a population transfer between the polariton branches \cite{Muller2015, Hohenester2009}. This allows us to monitor the coherent interaction between a laser and the polariton rung UP1 by monitoring the spectrally-filtered emission from LP1. Intuitively, one would expect this phonon-assisted emission to be weak. However, for a positive detuning of the QD, the radiative recombination rate $\Gamma_{\text{UP1}}^r$ of UP1 is strongly Purcell suppressed (as a result of the photonic bandgap), while the radiative decay rate $\Gamma_{\text{LP1}}^r$ of LP1 is very fast. Therefore, at detunings of $\Delta=5-10~g$, where the phonon-assisted population transfer rate from UP1 to LP1 $\Gamma_{\text{f}}^{nr}$ is large compared to $\Gamma_{\text{UP1}}^r$,  most emission actually occurs from LP1 \cite{Muller2015}.

In order to further reduce a leakage of the coherently scattered light into our detection channel, the technique of self-homodyne suppression (SHS) \cite{Fischer2015} is applied, where we carefully adjust polarization and focus to interfere the JC coherently scattered light with light scattered from above-the-light-line modes destructively. 

\subsection*{Rabi oscillations}

Rabi oscillations are a fundamental signature of the coherent interaction between the polaritons and the excitation laser. To this end, we perform power-dependent measurements, where we resonantly excite UP1 while we detect the emission from LP1 as discussed above. Note that we did not subtract a background from the data, since the laser light is spectrally separated from the emission of LP1 and thus is excluded by spectral filtering. A typical measurement for a QD-cavity detuning of $\Delta = 6.75~g$ with $18$~ps long pulses and increasing excitation power is presented in figure \ref{fig:General}c. Clearly, strongly damped oscillations are observed.

We drive the system resonantly as illustrated with a blue arrow in figure \ref{fig:General}a. In the Bloch sphere this can be described by an excitation pulse inducing a rotation of the Bloch vector about a rotation axis $x$. This pulse can generate superpositions of the ground and excited state. With increasing excitation power, the Bloch vector is brought all the way to the excited state and further to the ground state, again. A pulse with the area of $(m+1)\pi$, with $m= 0,1,2,...$ results in the excited state yielding maximum emission from the system, while a pulse with the area of $2m\pi$ results in the ground state yielding minimum emission. 

In order to get a better understanding of the damping mechanisms, we developed a quantum optical model based on a phenomenological two-level system using the Quantum Toolbox in PYTHON (QuTiP) \cite{Johansson2013}. A fit to the data with this model is shown in figure \ref{fig:General}c as a red line and produces good overall agreement. This simulation reveals that dephasing originating from phonons \cite{Ramsay2010a} and excitation pulses is causing the strong damping observed in the measurements.

\subsection*{Ramsey fringes}
\begin{figure}[ht]
	\centering
		\includegraphics{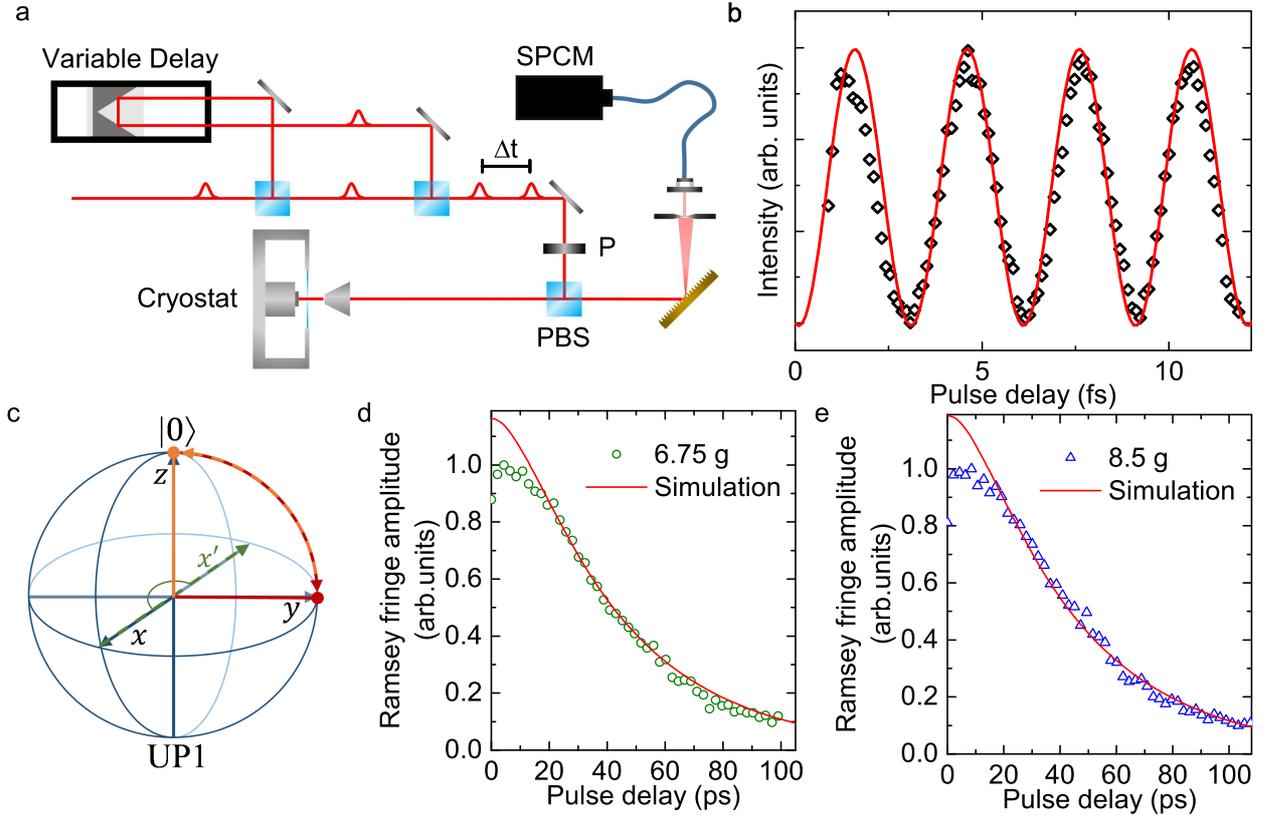}
	\caption{\textbf{Ramsey fringes of the strongly coupled QD-cavity system.} (\textbf{a}) Illustration of the setup for two-pulse experiments. The separation of the excitation pulses is tuned with a linear delay line in the ps- and a piezo-controlled retroreflector in the fs-regime. (\textbf{b}) Ramsey fringes obtained by detecting emission from LP1 while exciting UP1 with two $\frac{\pi}{2}$ pulses, separated by $\Delta t \approx 43$~ps and altered by $0-11$~fs. Simulations to the data are shown as a red line. (\textbf{c}) Illustration of the Bloch sphere with $|0\rangle$ as the ground state and UP1 as the excited state. Under two-pulse excitation, the Bloch vector gets rotated about the $x$ axis onto the equator by the first $\frac{\pi}{2}$-pulse (illustrated in red). The phase difference between the first and the second excitation pulse introduces a new rotation axis $x^\prime$ (green dashed line). Finally, the second $\frac{\pi}{2}$-pulse rotates the Bloch vector about the $x^\prime$ axis towards the ground state (illustrated in orange). Evolution of the amplitude of Ramsey fringes for QD-cavity detunings of (\textbf{d}) $\Delta = 6.75~g$ and (\textbf{e}) $\Delta = 8.5~g$ with increasing pulse separation. The red lines represent simulations to the data, based on measured lifetimes \cite{Muller2015} of $65 \pm 5$~ps and $71 \pm 7$~ps, respectively. The simulation is normalized to the maximum contrast of the experimental data. An exponential fit of the decay reveals dephasing times of $T_2^*\left[6.75~g\right]=72.09\pm 5$~ps and $T_2^*\left[8.5~g\right]=70.13 \pm 7$~ps.}
	\label{fig:Ramsey}
\end{figure}

To obtain the system's coherence properties, we performed two-pulse experiments. For that purpose, the excitation path was altered as illustrated in figure \ref{fig:Ramsey}a. A 50:50 beamsplitter divides the excitation pulse, sending one pulse through a linear delay line. This allows for variable pulse delays in the ps-regime, while fine tuning of the delay in the fs-regime is realized with a piezo-controlled retroreflector. Therefore, we are able to excite the system using two pulses with precise control of the inter-pulse time delay. 

From the Rabi oscillations in figure \ref{fig:General}c, the excitation power for each arm of the delay line is carefully chosen to be $\frac{\pi}{2}$. Typical results of such a two-pulse experiment are presented in figure \ref{fig:Ramsey}b. Here, the QD-cavity detuning is $\Delta = 6.75~g$, while the coarse pulse separation is set to be $\Delta t\approx 43$~ps and is altered in the range of $0-11$~fs. Clear oscillations - Ramsey fringes - are observed. The rotation of the Bloch vector is illustrated in figure \ref{fig:Ramsey}c. Starting in the ground state, the first $\frac{\pi}{2}$-pulse pivots the Bloch vector about the rotation axis $x$, resulting in a superposition lying on the equator of the Bloch sphere (illustrated in red in figure \ref{fig:Ramsey}c). In the rotating frame, the Bloch vector does not precess on the equator. Instead, the phase difference of the two excitation pulses introduces a second rotation axis $x^\prime$ (green dashed line in figure \ref{fig:Ramsey}c).  Finally, the second $\frac{\pi}{2}$-pulse pivots the Bloch vector about $x^\prime$, which is either towards the ground state (as shown in orange in figure \ref{fig:Ramsey}c) or towards the excited state, depending on the phase difference of the two pulses. Thus, the final state is strongly dependent on the pulse separation. This dependence can be monitored by detecting the emission from the excited state. After the arrival of the second excitation pulse, the system can end up in the excited state corresponding to the maxima in the Ramsey fringes, the ground state corresponding to the minima (as illustrated in figure \ref{fig:Ramsey}c), or states in between. 

To simulate this experiment, we take the measured polariton lifetime ($65 \pm 5$~ps at $\Delta = 6.75~g$) \cite{Muller2015}, the damping parameters obtained for the Rabi oscillations (presented above) and the alternated pulse separation into account. Then we calculate the population probability of the excited state. The results are presented in figure \ref{fig:Ramsey}b as a red line and show good agreement with the measured emission from LP1.

In order to experimentally determine the dephasing time $T_2^*$ of our system, we repeat the Ramsey fringe measurements for coarse pulse separations from $0$~ps to $>100$~ps. The evolution of the fringe amplitudes with increasing pulse separation for QD-cavity detunings of $\Delta = 6.75~g$ and $\Delta = 8.5~g$ are shown in figures \ref{fig:Ramsey}d and \ref{fig:Ramsey}e, respectively.

We extract the dephasing times $T_2^*$ by fitting the data with an exponential decay function, resulting in $T_2^*\left[6.75~g\right]=72.09 \pm 5$~ps and $T_2^*\left[8.5~g\right]=70.13 \pm 7$~ps, respectively. 

Using the same model as in figure \ref{fig:Ramsey}b and the experimentally obtained polariton lifetimes of $65 \pm 5$~ps ($\Delta = 6.75~g$) and $71 \pm 7$~ps ($\Delta = 8.5~g$)\cite{Muller2015}, we can reproduce the measured data in simulations. The results are presented as red solid lines in figures \ref{fig:Ramsey}d and \ref{fig:Ramsey}e, showing good agreement. Our simulations reveal that the dephasing time is mainly limited by an excitation power-dependent and a phonon-induced dephasing \cite{Ramsay2010b, Ramsay2010e}. Only for small delay times, the measured amplitudes are smaller than the simulated ones. The mismatch for small pulse separations between data and simulation may result from the interference of the excitation pulses when they overlap in time. Although the model takes this into account, it is not perfectly consistent with the data, probably due to the experimentally limited stability of the phase, which is extremely sensitive when the pulses overlap.

\subsection*{Complete Coherent Control}
We now turn our attention to complete coherent control of the polariton on the Bloch sphere. To this end, we perform experiments in which we vary both, the excitation power and phase difference in a two-pulse experiment.
\begin{figure}[ht]
	\centering
		\includegraphics{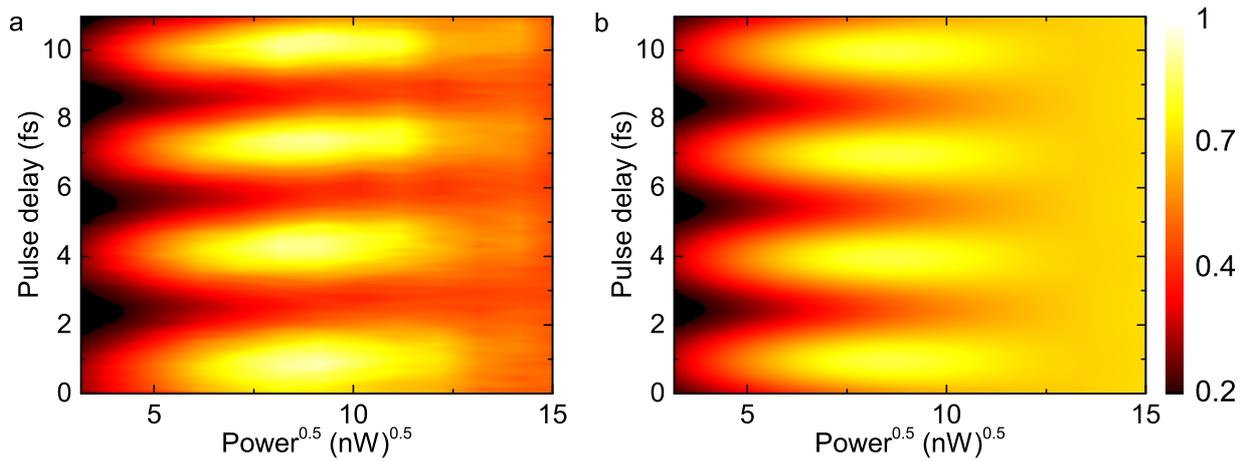}
		\caption{\textbf{Complete coherent control of the strongly coupled QD-photonic crystal cavity system.} (\textbf{a}) Excitation power and inter-pulse separation dependent measurement for a coarse pulse separation of $ 40$~ps. (\textbf{b}) Quantum optical simulation of complete coherent control, showing good agreement with the experimental data. The emission intensity of the data resulting from experiment and simulation share the same colorscale.}
	\label{fig:SU2}
\end{figure}

In figure \ref{fig:SU2}a we present the measured emission from LP1 under resonant excitation of UP1 with two pulses. The pulse separation in this measurement is set to $\approx 40$~ps and altered by $0-11$~fs, while the excitation pulse powers are independently increased. 

The data clearly reveals first-order maxima at an excitation power of $\frac{\pi}{2}$ per pulse, followed by minima at an excitation power of $\pi$ per pulse. We note here that resolving the second-order maxima with sufficient contrast is extremely difficult, since the strong power-dependent damping resulting from the polariton-phonon coupling increases with the laser power. This effect originates from the hybridization of the polariton levels. With increasing excitation power higher rungs start to mix with UP1, opening up channels for more phonon interaction within the system \cite{Muller2015}. Another effect reducing the visibility of the second-order maxima is a very small sample drift in the He flow cryostat during the long measurement time of $30$~min and longer. Due to this drift, the effect of the sensitive technique SHS is reduced. In addition to that, the excitation and the detection efficiency suffer from the drift. 

Again, we can explain the experiment in the Bloch sphere picture: Depending on the excitation power, the first pulse will rotate the Bloch vector to a corresponding superposition of ground and excited state. Thus, using the phase difference between the pulses, the second pulse can rotate the Bloch vector across the sphere and we can access any point on the Bloch sphere by varying the excitation power.

For our experimental results this means that complete coherent control is successfully demonstrated. The whole Bloch sphere can be accessed with an excitation power of $\pi$ per pulse, which corresponds to the first-order minima in the complete coherent control measurements. 

Our simulations with the quantum optical model that already show good agreement for Rabi oscillations and Ramsey fringes support our experimental findings and are presented in figure \ref{fig:SU2}b.

\section*{Conclusion}

In this work, we investigate the possibility to perform coherent optical control of a QD strongly coupled to a photonic crystal cavity. The strong exciton-phonon interaction within the system facilitates a population transfer from the upper to the lower polariton, which significantly increases the emission rates and allows for sophisticated excitation-detection schemes. Using a cross-polarized setup combined with spectral filtering and self-homodyne \cite{Fischer2015} suppression allows us to infer the population of the polariton state after the interaction with one or two short optical pulses.

With the observation of Rabi oscillations, we provide fundamental confirmation of the coherent interaction between the laser and a polaritonic state. Moreover, the successful demonstrations of Ramsey fringes and finally complete coherent control provide evidence that on-chip integrated photonic crystal cavity polariton-based quantum optical systems can be controlled. Compared to QDs alone, we can benefit from all advantages of the QD-photonic crystal cavity system, such as the efficient photonic interface \cite{Englund2005, Muller2015c, Hennessy2006, Laurent2005, Majumdar2012} and unprecedented fast emission rates in non-classical light generation \cite{Muller2015}, while retaining full access to the Bloch sphere.

To support, understand and interpret the experimental results, we also developed a phenomenological quantum optical model, which produces good overall agreement with the measurements. We find significant phonon-induced dephasing, which comes along with a population transfer from UP1 to LP1, facilitating single photon emission from LP1 \cite{Muller2015} and indistinguishable photon generation at elevated temperature and unprecedented fast emission rates from UP1 \cite{Muller2015c}. This interaction also has a strong influence on the excitation conditions for high-fidelity photon blockade \cite{Muller2015a, Muller2015}. In particular, the phonon interaction shortens the lifetime of the polaritons and shorter excitation pulse lengths are required to reach the lowest second order coherence values. Combining our demonstration of coherent control with efficient coupling to waveguides \cite{Faraon2007}, would make strongly coupled QD-photonic crystal cavity systems viable candidates for non-classical on-chip photonic devices.

\section*{Methods}
\subsection*{Sample fabrication}

The MBE-grown structure consists of an $\sim 900$~nm thick $\text{Al}_{0.8}\text{Ga}_{0.2}\text{As}$ sacrificial layer followed by a $145~$nm thick GaAs layer containing a single layer of InAs QDs. Our growth conditions result in a typical QD density of $(60 - 80)$~$\mu m^{-2}$. Using $100$~keV e-beam lithography with ZEP resist, followed by reactive ion etching and HF removal of the sacrificial layer, we define the photonic crystal cavity. The photonic crystal lattice constant was $a = 246$~nm and the hole radius $r \sim 60$~nm. The cavity fabricated is a linear three-hole defect (L3) cavity. To improve the cavity quality factor, holes adjacent to the cavity were shifted.
\subsection*{Optical spectroscopy}
All optical measurements were performed with a liquid helium flow cryostat at temperatures in the range $20-30$~K. For excitation and detection, a microscope objective with a numeric aperture of $\text{NA} = 0.75$ was used. Cross-polarized measurements were performed using a polarizing beam splitter. To further enhance the extinction ratio, additional thin film linear polarizers were placed in the excitation/detection pathways and a single mode fibre was used to spatially
filter the detection signal. Furthermore, two waveplates were placed between the beamsplitter and microscope objective: a half-wave plate to rotate the polarization relative to the cavity and a quarter-wave plate to correct for birefringence of the optics and sample itself. Photons are detected after spectral filtering with a single photon avalanche diode.

The thin film polarizers and polarizing beamsplitters allow us to achieve an extinction ratio of $10^{-7}$ between excitation and detection path on bulk. This suppression ratio is large enough such that light geometrically rotated by the high NA objective plays little role in the ultimate laser suppression. Instead, the amount of light classically scattered into the detection channel is determined by the fidelity of the self-homodyne effect. Experimentally, we previously found that this effect was capable of interferometrically cancelling $> 95~\%$ of the light scattered through the L3 cavity's fundamental mode \cite{Fischer2015}. In light of this strong suppression, no background has been subtracted from the experimental data.

Throughout the measurements we use a picosecond pulsed laser with $80.2$~MHz repetition rate with $3$-ps laser pulses. We use a 4f pulse shaper with an $1800$~lines/mm grating and a $40$~cm lens to create $18$~ps long pulses.
\subsection*{Simulations}
The simulations in this paper are performed using the Quantum Toolkit in PYTHON (QuTiP) \cite{Johansson2013}. The model contains of a two level system described by the Hamiltonian $\mathcal{H} = \mathcal{H}_0 + \mathcal{H}_D$, where $\mathcal{H}_0$
\begin{equation}
\mathcal{H}_0 = \omega_{cgs} |0\rangle \langle 0| + (\omega_0 - \omega_{cgs} - \omega_L) |1\rangle \langle1| 
\end{equation}
is the unperturbed term in the rotating frame with the ground state frequency $\omega_{cgs} = 0$, the eigenfrequency of the system $\omega_0$ and the frequency of the excitation laser $\omega_L = \omega_0$. While the driving term $\mathcal{H}_D$ is given by 
\begin{equation}
\mathcal{H}_D = \Omega \cdot (1+e^{i \omega_L \Delta t}) \cdot (|0\rangle \langle 1| + |1\rangle \langle 0|),
\end{equation}
with excitation power $\Omega$. The model includes excitation with either one excitation pulse for Rabi oscillations or with two laser pulses with a pulse separation of $\Delta t$, introducing a phase shift, for Ramsey fringes and complete coherent control. 
For all experiments the evolution of the population of the excited state is calculated with respect to the time. In order to do so, the evolution of the density matrix is calculated in the rotating frame
\begin{equation}
\frac{d \tilde{\rho} \left(t\right)}{dt} =  - i \left[\mathcal{H},\tilde{\rho}\left(t \to \infty \right)\right] + \sum_j \mathcal{L}\left(c_j\right),
\label{eq:densitymatrix}
\end{equation} 

and we numerically integrated equation \ref{eq:densitymatrix}. Each collapse operator $c_j$ is included in the model as a Lindblad superoperator $\mathcal{L}\left(c_j\right)$. The collapse operators used are: The radiative decay
\begin{equation}
|0\rangle \langle1| \sqrt{\gamma_r},
\end{equation} 
with the detuning dependent radiative decay rate $\gamma$, the phonon dephasing 
\begin{equation}
|1\rangle \langle1| \sqrt{\gamma_{ph}},
\end{equation}
with a phonon dephasing rate of $\gamma_{ph} = \frac{1}{27~\text{ps}}$, and a dephasing term that depends on the excitation power
\begin{equation}
|1\rangle \langle1| \sqrt{\alpha_{P}} \Omega \left(e^{-\frac{\left(t-4\sigma\right)^2}{2\sigma^2}} + e^{-\frac{\left(t-\Delta t-4\sigma\right)^2}{2\sigma^2}}\right),
\end{equation}
with the fitting parameter $\alpha = \frac{1}{0.1}$.
This sum of population directly corresponds to the photon emission of the system \cite{Kaer2012} and produces good agreement to all experiments presented throughout this paper.


\begin{thebibliography}{10}

\bibitem{Monroe2002}
Monroe, C.,
\newblock {Quantum information processing with atoms and photons}.
\newblock {\em Nature} \textbf{416}, 238--246 (2002).

\bibitem{Vuckovic2003}
Vu\v{c}kovi\'{c}, J. \& Yamamoto, Y.,
\newblock {Photonic crystal microcavities for cavity quantum electrodynamics with a single quantum dot}.
\newblock {\em Appl. Phys. Lett.} \textbf{82}, 2374 (2003).

\bibitem{Jelezko2006}
Jelezko, F. \& Wachtrup, J.,
\newblock {Single defect centres in diamond: A review}.
\newblock {\em Phys. Status Solidi (A)} \textbf{203}, 3207--3225 (2006).

\bibitem{Michler2009}
Michler, P.,
\newblock {Single Semiconductor Quantum Dots}.
\newblock {\em Springer}, (2009).

\bibitem{MasBalleste2011}
Mas-Balleste, R., Gomez-Navarro, C., Gomez-Herrero, J. \& Zamora, F.,
\newblock {2D materials: to graphene and beyond}.
\newblock {\em R. Soc. Chem. Adv.} \textbf{3}, 20--30 (2011).

\bibitem{Stievater2001}
Stievater, T.H., \textit{et al.},
\newblock {Rabi Oscillations of Excitons in Single Quantum Dots}.
\newblock {\em Phys. Rev. Lett.} \textbf{87}, 133603 (2001).

\bibitem{Zrenner2002}
Zrenner, A., \textit{et al.},
\newblock {Coherent properties of a two-level system based on a quantum-dot
  photodiode.}
\newblock {\em Nature} \textbf{418}, 612--614 (2002).

\bibitem{Ramsay2010a}
Ramsay, A.J.,
\newblock {A review of the coherent optical control of the exciton and spin
  states of semiconductor quantum dots}.
\newblock {\em Semicond. Sci. Technol.} \textbf{25}, 103001 (2010).

\bibitem{Englund2005}
Englund, D., \textit{et al.},
\newblock {Controlling the Spontaneous Emission Rate of Single Quantum Dots in
  a 2D Photonic Crystal}.
\newblock {\em Phys. Rev. Lett.} \textbf{95}, 013904 (2005).

\bibitem{Brossard2010}
Brossard, F.S.F., \textit{et al.},
\newblock {Strongly coupled single quantum dot in a photonic crystal waveguide
  cavity}.
\newblock {\em Appl. Phys. Lett.} \textbf{97}, 111101 (2010).

\bibitem{Reinhard2011}
Reinhard, A., \textit{et al.},
\newblock {Strongly correlated photons on a chip}.
\newblock {\em Nature Photon.} \textbf{6}, 93--96 (2011).

\bibitem{Kim2013}
Kim, H., Bose, R., Shen, T., Solomon, G. \& Waks, E.,
\newblock {A quantum logic gate between a solid-state quantum bit and a
  photon}.
\newblock {\em Nature Photon.} \textbf{7}, 373--377 (2013).

\bibitem{Lodahl2004}
Lodahl, P., \textit{et al.},
\newblock {Controlling the dynamics of spontaneous emission from quantum dots
  by photonic crystals}.
\newblock {\em Nature} \textbf{430}, 654--657 (2004).

\bibitem{Wilson-Rae2001}
Wilson-Rae, I. \& Imamoglu, A.,
\newblock {Quantum dot cavity-QED in the presence of strong electron-phonon
  interactions}.
\newblock {\em Phys. Rev. B} \textbf{65}, 235311 (2002).

\bibitem{Muller2015c}
M\"{u}ller, K., \textit{et al.},
\newblock {Nanocavity-enabled ultrafast generation of highly-indistinguishable
  photons}.
\newblock {\em arXiv preprint}, arXiv:1512.05626 (2015).

\bibitem{Muller2015a}
M\"{u}ller, K., \textit{et al.},
\newblock {Coherent Generation of Nonclassical Light on Chip via Detuned Photon
  Blockade}.
\newblock {\em Phys. Rev. Lett.} \textbf{114}, 233601 (2015).

\bibitem{Muller2015}
M\"{u}ller, K., \textit{et al.},
\newblock {Ultrafast Polariton-Phonon Dynamics of Strongly Coupled Quantum
  Dot-Nanocavity Systems}.
\newblock {\em Phys. Rev. X} \textbf{5}, 031006 (2015).

\bibitem{Sun2015}
Sun, S., Kim, H., Solomon, G.S. \& Waks, E.,
\newblock {A quantum phase switch between a single solid-state spin and a photon}.
\newblock {\em Nat. Nano.}, DOI:10.1038/nnano.2015.334 (2016).

\bibitem{Blattmann2014}
Blattmann, R., Krenner, H.J., Kohler, S. \& H\"{a}nggi,  P.,
\newblock {Entanglement creation in a quantum-dot-nanocavity system by
  Fourier-synthesized acoustic pulses}.
\newblock {\em Phys. Rev. A} \textbf{89}, 012327 (2014).

\bibitem{Clark2007}
Clark, S.M., Fu, K.M.C., Ladd, T.D. \& Yamamoto, Y.,
\newblock {Quantum computers based on electron spins controlled by ultrafast
  off-resonant single optical pulses}.
\newblock {\em Phys. Rev. Lett.} \textbf{99}, 040501 (2007).

\bibitem{Akahane2003}
Akahane, Y., Asano, T., Song, B.-S.S. \& Noda, S.,
\newblock {High-Q photonic nanocavity in a two-dimensional photonic crystal}.
\newblock {\em Nature} \textbf{425}, 944--947 (2003).

\bibitem{Arakawa2012}
Arakawa, Y., Iwamoto, S., Nomura, M., Tandaechanurat, A. \& Ota, Y.,
\newblock {Cavity Quantum Electrodynamics and Lasing Oscillation in Single
  Quantum Dot-Photonic Crystal Nanocavity Coupled Systems}.
\newblock {\em IEEE J. Sel. Top. Quantum Electron.} \textbf{18}, 1818--1829 (2012).

\bibitem{Hennessy2006}
Hennessy, K., \textit{et al.},
\newblock {Quantum nature of a strongly coupled single quantum dot-cavity
  system}.
\newblock {\em Nature} \textbf{445}, 896--899 (2007).

\bibitem{Jaynes1963}
Jaynes, E.T. \& Cummings, F.W.,
\newblock {Comparison of quantum and semiclassical radiation theories with
  application to the beam maser}.
\newblock {\em Proc. IEEE} \textbf{51}, 89--109 (1963).

\bibitem{Laussy2014}
Laussy, F.P., del Valle, E., Schrapp, M., Laucht, A. \& Finley, J.J.,
\newblock {Climbing the Jaynes-Cummings ladder by photon counting}.
\newblock {\em J. Nanophotonics} \textbf{6}, 061803 (2012).

\bibitem{Reithmaier2004}
Reithmaier, J.P., \textit{et al.},
\newblock {Strong coupling in a single quantum dot-semiconductor microcavity
  system}.
\newblock {\em Nature} \textbf{432}, 197--200 (2004).

\bibitem{Peter2005}
Peter, E., \textit{et al.},
\newblock {Exciton-Photon Strong-Coupling Regime for a Single Quantum Dot
  Embedded in a Microcavity}.
\newblock {\em Phys. Rev. Lett.} \textbf{95}, 067401 (2005).

\bibitem{Englund2007}
Englund, D., \textit{et al.},
\newblock {Controlling cavity reflectivity with a single quantum dot}.
\newblock {\em Nature} \textbf{450}, 857--861 (2007).

\bibitem{Kaniber2008a}
Kaniber, M., \textit{et al.},
\newblock {Tunable single quantum dot nanocavities for cavity QED experiments}.
\newblock {\em J. Phys. Condens. Matter} \textbf{20}, 454209 (2008).

\bibitem{Kaniber2008}
Kaniber, M., \textit{et al.},
\newblock {Highly efficient single-photon emission from single quantum dots
  within a two-dimensional photonic band-gap}.
\newblock {\em Phys. Rev. B} \textbf{77}, 073312 (2008).

\bibitem{Gaudreau2011}
Gaudreau, L., \textit{et al.},
\newblock {Coherent control of three-spin states in a triple quantum dot}.
\newblock {\em Nat. Phys.} \textbf{8}, 54--58 (2011).

\bibitem{Webster2013}
Webster, L.A., \textit{et al.},
\newblock {Coherent Control to Prepare an InAs Quantum Dot for Spin-Photon
  Entanglement}.
\newblock {\em Phys. Rev. Lett.} \textbf{112}, 126801 (2014).

\bibitem{Dominici2014}
Dominici, L., \textit{et al.},
\newblock {Ultrafast Control and Rabi Oscillations of Polaritons}.
\newblock {\em Phys. Rev. Lett.} \textbf{113}, 226401 (2014).

\bibitem{Fischer2015}
Fischer, K.A., \textit{et al.},
\newblock {Self-homodyne measurement of a dynamic Mollow triplet in the solid
  state}.
\newblock {\em Nature Photon.}, DOI:10.1038/nphoton.2015.276 (2016).

\bibitem{Ota2009}
Ota, Y., Iwamoto, S., Kumagai, N. \& Arakawa, Y.,
\newblock {Impact of interactions on quantum-dot cavity quantum
  electrodynamics}.
\newblock {\em arXiv preprint}, arXiv:0908.0788 (2009).

\bibitem{Hohenester2009}
Hohenester, U., \textit{et al.},
\newblock {Phonon-assisted transitions from quantum dot excitons to cavity
  photons}.
\newblock {\em Phys. Rev. B} \textbf{80}, 201311 (2009).

\bibitem{Johansson2013}
Johansson, J.R., Nation, P.D. \& Nori, F.,
\newblock {QuTiP 2: A Python framework for the dynamics of open quantum
  systems}.
\newblock {\em Comput. Phys. Commun.} \textbf{184}, 1234--1240 (2013).

\bibitem{Ramsay2010b}
Ramsay, A.J., \textit{et al.},
\newblock {Damping of Exciton Rabi Rotations by Acoustic Phonons in Optically Excited InGaAs/GaAs Quantum Dots}.
\newblock {\em Phys. Rev. Lett.} \textbf{104}, 017402 (2010).

\bibitem{Ramsay2010e}
Ramsay, A.J., \textit{et al.},
\newblock {Phonon-Induced Rabi-Frequency Renormalization of Optically Driven Single InGaAs/GaAs Quantum Dots}.
\newblock {\em Phys. Rev. Lett.} \textbf{105}, 177402 (2010).

\bibitem{Laurent2005}
Laurent, S., \textit{et al.},
\newblock {Indistinguishable single photons from a single-quantum dot in a
  two-dimensional photonic crystal cavity}.
\newblock {\em Appl. Phys. Lett.} \textbf{87}, 163107 (2005).

\bibitem{Majumdar2012}
Majumdar, A., Rundquist, A., Bajcsy, M. \& Vu\v{c}kovi\'{c}, J.,
\newblock {Cavity quantum electrodynamics with a single quantum dot coupled to
  a photonic molecule}.
\newblock {\em Phys. Rev. B} \textbf{86}, 045315 (2012).

\bibitem{Faraon2007}
Faraon, A., Waks, E., Englund, D., Fushman, I. \& Vu\v{c}kovi\'{c}, J.,
\newblock {Efficient photonic crystal cavity-waveguide couplers }.
\newblock {\em Appl. Phys. Lett.} \textbf{90}, 073102 (2007).

\bibitem{Kaer2012}
Kaer, P., Nielsen, T.R.R., Lodahl, P., Jauho, A.P. \& M\o~ rk, J.,
\newblock {Microscopic theory of phonon-induced effects on semiconductor
  quantum dot decay dynamics in cavity QED}.
\newblock {\em Phys. Rev. B} \textbf{86}, 085302 (2012).


\end{thebibliography}

\noindent
\section*{Acknowledgements}
We gratefully acknowledge financial support from the Air Force Office of Scientific Research, MURI center for multifunctional light-matter interfaces based on atoms and solids (Grant No. FA9550-12-1-0025) and support
from the Army Research Office (Grant No. W911NF1310309). KAF acknowledges support from the Lu Stanford Graduate Fellowship and the National Defense Science and Engineering Graduate Fellowship. KM acknowledges support from the Alexander von Humboldt Foundation. KGL acknowledges support from the Swiss National Science Foundation. JLZ acknowledges support from the Stanford  Graduate Fellowship. YK acknowledges support from the Stanford Graduate Fellowship and the National Defense Science and Engineering Graduate Fellowship.
\section*{Author contributions statement}
CD, KM and KGL performed the experiments. TS performed the MBE growth of the QD structure. AR and TS fabricated the photonic crystal device. CD and KAF performed the theoretical work and simulations. JLZ and YK provided expertise. JV supervised the entire project. All authors contributed to discussions and writing the manuscript.

\section*{Competing financial interests}
The authors declare no competing financial interests.

\end{document}